\documentclass[sigconf]{acmart}
\usepackage{enumitem}
\usepackage{multirow}
\usepackage{subfigure}
\usepackage{blindtext}
\usepackage{enumitem}
\usepackage{algorithm}  
\usepackage{algpseudocode}  
\usepackage{amsmath}  
\usepackage{hyperref}
\AtBeginDocument{%
  }

\setcopyright{acmcopyright}
\copyrightyear{2023}
\acmYear{2023}
\setcopyright{acmlicensed}
\acmConference[SIGIR '23] {Proceedings of the 46th International ACM SIGIR Conference on Research and Development in Information Retrieval}{July 23--27, 2023}{Taipei, Taiwan.}
\acmBooktitle{Proceedings of the 46th International ACM SIGIR Conference on Research and Development in Information Retrieval (SIGIR '23), July 23--27, 2023, Taipei, Taiwan}
\begin{document}

% \title{Policy Constrained Offline Q-learning with Mixed Dataset for Recommender Systems }

\title{MDDL: A Framework for Reinforcement Learning-based Position Allocation in Multi-Channel Feed}


\author{Xiaowen Shi}
\authornotemark[1]
\affiliation{%
 \institution{Meituan}
 \city{Beijing}
 \country{China}
}
\email{shixiaowen03@meituan.com}

\author{Ze Wang}
\authornote{Equal contribution. Listing order is random.}

\authornote{Corresponding author.}
\affiliation{%
 \institution{Meituan}
 \city{Beijing}
 \country{China}
}
\email{wangze18@meituan.com}

\author{Yuanying Cai}
\authornote{This work was done when Yuanying Cai was an intern in Meituan.}
\affiliation{%
 \institution{IIIS, Tsinghua University}
 \city{Beijing}
 \country{China}
}
\email{yuanying.cc@gmail.com}

\author{Xiaoxu Wu}
\affiliation{%
 \institution{Meituan}
 \city{Beijing}
 \country{China}
}
\email{wuxiaoxu04@meituan.com}

\author{Fan Yang}
\affiliation{%
 \institution{Meituan}
 \city{Beijing}
 \country{China}
}
\email{yangfan129@meituan.com}

\author{Guogang Liao}
\affiliation{%
 \institution{Meituan}
 \city{Beijing}
 \country{China}
}
\email{liaoguogang@meituan.com}

\author{Yongkang Wang}
\affiliation{%
 \institution{Meituan}
 \city{Beijing}
 \country{China}
}
\email{wangyongkang03@meituan.com}

\author{Xingxing Wang}
\affiliation{%
 \institution{Meituan}
 \city{Beijing}
 \country{China}
}
\email{wangxingxing04@meituan.com}

\author{Dong Wang}
\affiliation{%
 \institution{Meituan}
 \city{Beijing}
 \country{China}
}
\email{wangdong07@meituan.com}
\renewcommand{\shortauthors}{Xiaowen Shi and Ze Wang, et al.}
%%
%%
%% Sometimes the addresses are too long to fit on the page.  In this
%% case uncomment the lines below and fill them accodingly.
%%
%% \authorsaddresses{Corresponding author: Ben Trovato,
%% \href{mailto:trovato@corporation.com}{trovato@corporation.com};
%% Institute for Clarity in Documentation, P.O. Box 1212, Dublin,
%% Ohio, USA, 43017-6221}

\begin{abstract}
% In industrial scenarios, the policy models of the recommendations system are usually only allowed to be trained offline due to security concerns and directly applying reinforcement learning algorithms may result in the out-of-distribution problem and harms the performance. To alleviate the problem, we usually train the models with the uniformly distributed data collected by the random policy. However, we can only obtain a small number such data since rolling out the random policy model online with a large scale are not feasible, which results in the overfitting problem of the trained model. We further observe directly adding the large amount of dataset collected by the online running policy models with guaranteed performance to solve the overfitting problem can also harm the performance due to the OOD problem. To better use the large amount of datasets, we propose XX algorithm that use the large amount of datasets to provide imitation learning signals with a novel design to impose the imitation learning signal for the discrete control tasks. In the experiments, we evaluate our method in the integrated recommendation system on Meituan feed and demonstrate that our method significantly outperforms the previous baseline (CrossDQN). For now, our model has been fully deployed on Meituan feed and serves for more than 300 million users.

Nowadays, the mainstream approach in position allocation system is to utilize a reinforcement learning model to allocate appropriate locations for items in various channels and then mix them into the feed. 
There are two types of data employed to train reinforcement learning (RL) model for position allocation, named strategy data and random data. 
Strategy data is collected from the current online model, it suffers from an imbalanced distribution of state-action pairs, resulting in severe overestimation problems during training. 
On the other hand, random data offers a more uniform distribution of state-action pairs, but is challenging to obtain in industrial scenarios as it could negatively impact platform revenue and user experience due to random exploration.
As the two types of data have different distributions, designing an effective strategy to leverage both types of data to enhance the efficacy of the RL model training has become a highly challenging problem.

In this study, we propose a framework named \textbf{M}ulti-\textbf{D}istribution \textbf{D}ata \textbf{L}earning (\textbf{MDDL}) to address the challenge of effectively utilizing both strategy and random data for training RL models on mixed multi-distribution data. Specifically, MDDL incorporates a novel imitation learning signal to mitigate overestimation problems in strategy data and maximizes the RL signal for random data to facilitate effective learning. 
In our experiments, we evaluated the proposed MDDL framework in a real-world position allocation system and demonstrated its superior performance compared to the previous baseline. MDDL has been fully deployed on the Meituan food delivery platform and currently serves over 300 million users.
\end{abstract}

\keywords{Reinforcement Learning; Multi-Distribution Data Learning; Position
Allocation}

\maketitle

\section{Introduction}
\begin{figure}[tb]
  \centering
  \includegraphics[width=1\linewidth]{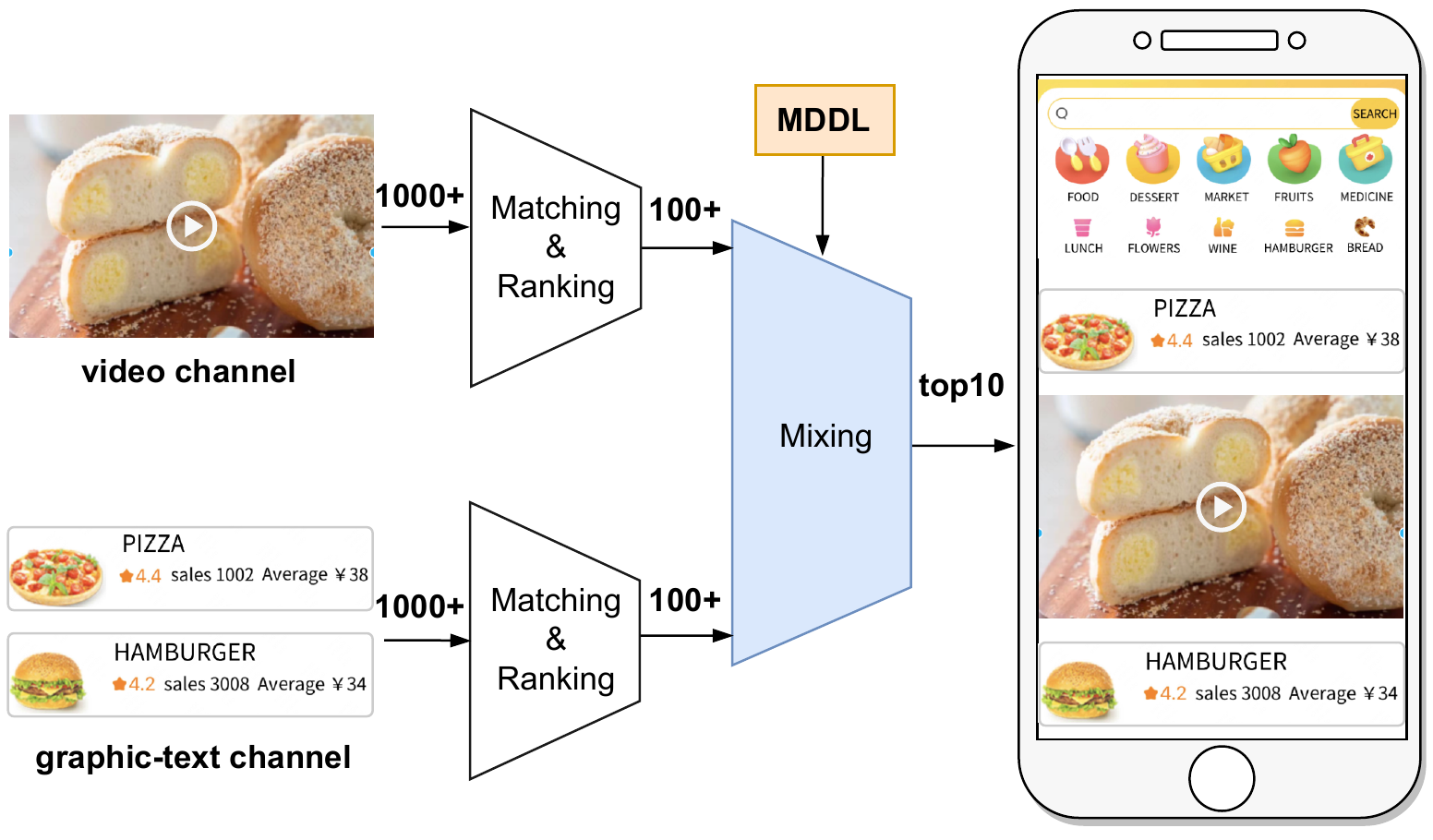}
  \caption{
    Structure of the position allocation system on Meituan food delivery platform.
  }
  \label{fig:fig1}
  \vspace{-0.2in}
\end{figure}

In the e-commerce scenario, a diverse array of items from different channels are aggregated and presented to users in feed \cite{xia2022balancing,wang2022off}. 
As shown in Figure \ref{fig:fig1}, the Meituan food delivery platform mainly provides two formats for presenting items, specifically graphic-text and video formats. 
Determining an optimal allocation of graphic-text and video positions that accommodates users' individual preferences while maximizing platform benefits has emerged as a critical research question. 
There exist two primary strategies for determining the display positions of items, namely the fixed position strategy \cite{chen2022extr,li2020deep,ouyang2020minet} and the dynamic position strategy \cite{xie2021hierarchical,zhao2019deep,liao2022crossdqn,wang2022learning,liao2022deep,wang2022hybrid,zhao2020jointly}.
The fixed position strategy involves allocating video items to predetermined positions. This strategy fails to account for personalized user information and is thus suboptimal. 
The dynamic allocation strategy, on the other hand, approaches the problem as a Markov Decision Process (MDP) \cite{sutton1998introduction}, which is subsequently solved via reinforcement learning (RL) \cite{liao2022crossdqn,xie2021hierarchical,zhao2020jointly}. For instance, \citet{xie2021hierarchical} propose a hierarchical RL framework that first determines the channel for each position before selecting a specific item to display. \citet{liao2022crossdqn} propose CrossDQN, which treats the combination of positions as an action in order to determine the allocation of products within a single screen at a time. This approach fully models the mutual influence between adjacent items.

\begin{figure}[tb]
  \centering
  \includegraphics[width=0.8\linewidth]{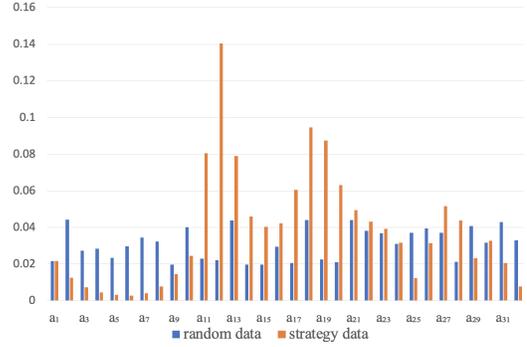}
  \caption{
  Distribution of actions on a certain type of state.
  }
  \label{fig:fig2}
  
\end{figure}

Previous research has predominantly centered around the design of reinforcement learning models to achieve superior performance. However, the quality of data is equally significant for reinforcement learning, yet has not been well explored in prior works. Two types of data can generally be utilized for training reinforcement learning models. The first is called strategy data, which comprises data collected by the state-of-the-art model that operates online. Although strategy data is adequate and enables the utilization of various features, the distribution of state-action pairs is highly imbalanced (see Figure \ref{fig:fig2}). For instance, samples collected from users who prefer videos are mostly of high video proportions. This skewed distribution can result in severe overestimation problems during the reinforcement learning training process\cite{kumar2020conservative}.
Particularly when utilizing advanced models like CrossDQN, the overestimation problems are further exacerbated by the high-dimensional state-action space\cite{van2016deep}. Even when employing techniques such as DoubleDQN \cite{van2016deep}, MaxminDQN \cite{lan2020maxmin}, and other methods \cite{meng2021effect, cini2020deep, chen2021randomized} to mitigate overestimation, it is still difficult to obtain a policy with better performance. The second type of data is random data collected through a random strategy. While the distribution of state-action pairs in random data is relatively uniform, the random strategy can have a detrimental impact on platform benefits. Moreover, random data can only be obtained through limited traffic in industrial scenarios, thereby constraining the utilization of effective features such as sparse ID features, which also hinders the performance of the RL model. Therefore, determining how to effectively utilize both types of data to enhance the efficacy of the RL model remains a highly challenging problem.

To this end, we propose a framework named \textbf{M}ulti-\textbf{D}istribution \textbf{D}ata \textbf{L}earning (\textbf{MDDL}), which is able to effectively utilize two kinds of learning signals to train RL agents on mixed multi-distribution data.
For strategy data, which is collected from the approximate expert strategy that runs well online, we impose the novel imitation learning signal \cite{osa2018algorithmic} to reduce the impact of overestimation problems.
For random data, we make full use of the RL signal to enable agent to learn effectively. While this data has a relatively uniform distribution of state-action pairs and is less prone to overestimation problems, it is limited in terms of its ability to fully utilize effective features such as sparse ID features.
By incorporating both signals, MDDL is able to strike a balance between exploration and exploitation, enabling the RL agent to learn more effectively from both types of data. Our proposed framework has the potential to improve the performance of RL agents in industrial scenarios where large amounts of mixed multi-distribution data are available.

The contribution of our work can be summarized as follows:
\begin{itemize}[leftmargin=*]
    \item \textbf{A novel multi-distribution data learning framework}. The MDDL framework we proposed can effectively learn on multi-distribution data and  perform better for position allocation problems. To the best of our knowledge, this is the first multi-distribution data learning framework for industrial position allocation.
    \item \textbf{Complete evaluation indicators for overestimation}. We provides a comprehensive approach to evaluate and locate overestimation problems in RL model training for industrial scenarios.
    \item \textbf{Detailed industrial experiments}. Our method significantly outperforms the all baselines in both offline experiments and online A/B tests. For now, our framework has been fully employed on Meituan food delivery platform. 
\end{itemize}

\section{PROBLEM FORMULATION}

In our scenario, we mix items from video and graphic-text channel\footnote{We set the number of channels to 2 for ease of illustration, but our method can be readily extended to scenarios where the number of channels is larger than 2.} and present $K$ slots in one screen and handle the position allocation for each screen in the feed of a request sequentially.
The position allocation problem is formulated as a Markov Decision Process (MDP) \cite{sutton1998introduction}  ($\mathcal{S}$, $\mathcal{A}$, $r$, ${P}$, $\gamma$), the elements of which are defined as follows:

\begin{itemize}[leftmargin=*]
\item \textbf{State space $\mathcal{S}$}. A state $s \in \mathcal{S}$ contains various information about candidate items available at step $t$ from the video channel and the graphic-text channel , such as their IDs. 
\item \textbf{Action space $\mathcal{A}$}. An action $a \in \mathcal{A}$ is the decision of displaying a video or a graphic-text at each position on the current screen, which is formulated as follows:
\begin{equation}
    a=(x_{1}, x_{2}, \ldots, x_{K}),
\end{equation}
where $x_{k}=\begin{cases}
    1 &\text{display a video in the $k$-th position} \\ 
    0 &\text{otherwise} 
    \end{cases}$, $\forall k \in [K]$.

\item \textbf{Reward $r$}. After the system takes an action in one state, a user browses the mixed list and gives a feedback. We use the Gross Merchandise Volume (GMV) on the screen as $r$. 
\item \textbf{Transition probability ${P}$}.
$P$ is the state transition probability from $s_t$ to $s_{t+1}$ after taking the action $a_t$. 
When user pulls down, the state $s_{t}$ transits to the state of next screen $s_{t+1}$. 
The items selected to present by $a_t$ will be removed from the state on next step $s_{t+1}$.
If user no longer pulls down, the transition terminates. 

\item \textbf{Discount factor $\gamma$}. The discount factor $\gamma \in [0, 1]$ balances the short-term and long-term rewards.

\end{itemize}
We use $\mathcal{D}_{m}=\{(s^t,a^t,r^t,s^{t+1})\}$ to denote the data collected by model policy (hereinafter referred to as strategy data) and use $\mathcal{D}_{r}=\{(s^t,a^t,r^t,s^{t+1})\}$ to denote the data collected by random policy (hereinafter referred to as random data). 
Given the MDP formulated as above, the objective is to find a position allocation policy $\pi: \mathcal{S} \rightarrow \mathcal{A}$ on $\mathcal{D}_{m} \cup \mathcal{D}_{r}$ to maximize the cumulative reward.

\section{METHODOLOGY}

In this section, we will detail how to use data collected by model policy and random policy to train reinforcement learning strategies in industrial scenarios.
To train on $\mathcal{D}_{m}$, we define the similarity between different actions using the weighted exposure ratio (WER) and introduce the imitation learning signals to help the agent learn. For random data $\mathcal{D}_{r}$, we train the agent using reinforcement learning signals based on the Bellman equation. 
Next, we will introduce the concept of WER first and then explain how to use the two signals for training.

\subsection{Weighted Exposure Ratio (WER)}
The task of imposing an imitation learning signal in offline reinforcement learning (RL) requires defining the similarity between actions. However, most recent methods that incorporate imitation learning signals in offline RL are designed for continuous control tasks \cite{osa2018algorithmic, hua2021learning, ho2016generative}, where the $\ell_2$ distances between actions are sufficient to measure similarity.
In contrast, our work focuses on the position allocation problem, where the action space is discrete. This presents a new challenge in defining the imitation learning signal, as it is not reasonable to directly calculate the distance between two actions represented by 0-1 vectors.
As an illustration, consider three different actions: $a_1=(1, 1, 1, 0, 0)$, $a_2=(0, 0, 1, 0, 0)$, and $a_3=(1, 1, 1, 1, 1)$. The distance between $a_1$ and $a_2$ is the same as the distance between $a_1$ and $a_3$, while the distance between $a_2$ and $a_3$ is far greater. The main reason why it is unreasonable to directly calculate the distance between actions is that it fails to consider the differences in the position and quantity of videos being exposed.

To overcome this challenge, we propose a new concept called the weighted exposure ratio (WER) to measure the similarity between actions. The WER converts each action into a specific number, which can be used to calculate the distance between different actions. Specifically, the WER of action $a_i^t$ at step $t$ is calculated as follows:

\begin{equation}
    \text{WER}(a_i^t) = \sum_{k=1,j={t\times K + k}}^{k=K} p_{j} \times \text{CTR}_j \times I_{i;k},
\end{equation}
where $p_{j}$ is the exposure probability for the $j$-th position in feed, $\text{CTR}_j$ it the click-through rate for the $j$-th position in feed, and $I_{i;k}$ is an indicator where $I_{i;k}=1$ means a video is inserted into the $k$-th position in current action $a_i^t$. 

\subsection{Model Training}
\subsubsection{Strategy Data}
As for strategy data, we use imitation learning for training. The learning goal is to make the action selected by the training model as consistent as possible with the original action in the sample.
Given a batch of transitions $B$, the loss for imitation learning is calculated as follows:

\begin{equation}
    L_\text{IL} \!  =\!   \frac{1}{|B|} \!  \! \sum_{(s^t,a^t,r^t,s^{t+1}\! )\in B} \! \! \! \Big( \text{WER}(a^t) - \text{WER}(\text{arg}\max_{a\in\mathcal{A}} Q(s^t,a)) \Big)^2.
\end{equation}
However, the argmax function is not differentiable.
Therefore, we use a soft version of argmax instead, i.e., we use 
\begin{equation}
  \text{WER}(\text{arg}\max_{a\in\mathcal{A}} Q(s^t,a))  \approx
  \sum_{i=1}^{N} \frac{1}{Z} \exp \big[\beta Q(s^t,a_i) \big] \text{WER}(a_i),
\end{equation}
where $N$ is the size of $\mathcal{A}$, $Z=\sum_{j=1}^{N} \exp [\beta Q(s^t,a_j) ]$ is the normalization factor and $\beta$ is the temperature coefficient. The larger the $\beta$ is, the closer the result is to argmax. 

\subsubsection{Random Data}
As for random data, we use the RL signal based on the Bellman equation \cite{mnih2015human}, calculated as follows:
\begin{equation}
  L_\text{RL} \! =\!  \frac{1}{|B|} \!  \sum_{(s^t,a^t,r^t,s^{t+1}\! )\in B} \! \! \Big( r^t\!  +\!  \gamma \max_{a'\in\mathcal{A}} Q(s^{t+1}, a') - Q(s^t,a^t) \Big) ^ 2.
\end{equation} 

\subsubsection{Offline Training}
For each iteration, we sample a batch of transitions from $\mathcal{D}_{m} \cup \mathcal{D}_{r}$, and update the parameter through the following loss:
\begin{equation}
  \label{eq:loss}
  \begin{aligned}
  &L = \alpha_1 \cdot L_\text{RL} + \alpha_2 \cdot L_\text{IL}, \\
      \text{where}\ \ &\begin{cases}
        \alpha_1=0, \alpha_2=1, & \forall (s^t,a^t,r^t,s^{t+1})\in \mathcal{D}_{m} \\ 
        \alpha_1=1, \alpha_2=0, &  \forall (s^t,a^t,r^t,s^{t+1})\in \mathcal{D}_{r} 
        \end{cases}.\end{aligned}
\end{equation}
where $\alpha_1$ and $\alpha_2$ are the coefficients to balance the two losses on mixed dataset $\mathcal{D}_{m} \cup \mathcal{D}_{r}$.

% This document provides \LaTeX\ templates for the article.

\section{Experiments}

In this section, we will verify the effect of our proposed method through extensive offline experiments and online A/B test. In offline experiments, we use crossDQN as the baseline model to compare the model performance and the degree of overestimation on different types of data, and analyze the influence of different hyperparameters. In online A/B test, we will compare our proposed method with the previous strategy deployed on the Meituan platform.

\subsection{Experimental Settings}

\subsubsection{Dataset}

We collect the dataset on the Meituan platform during October 2022. We use the data from 20221010 to 20221023 as the training set $\mathcal{D}_{train}$ and the data from 20221024 to 20221030 as the testing set $\mathcal{D}_{test}$. It is worth noting that the strategy data is collected from the online model which is trained through random data during September. We present the detailed statistics of the dataset in Table 1. In order to better reflect the distribution of the dataset, on the one hand, we convert the state representation into two dimensions through the t-SNE method, and cluster into 10 categories through the K-means. There are 320 types of state-action in total.

\begin{table}[bp]
  \caption{Statistics of the dataset. \ \ }
  \renewcommand\arraystretch{1.1}
  \centering
  \setlength{\tabcolsep}{1.4mm}{
  \begin{tabular}{c|cccc}
    \hline
  Data Type & \#Requests  &  \#Users  & \#Graphic-Texts & \# Videos \\
  \hline
  \hline
  Random & 13,235,289  &2,424,103   &  562,269 &  221,348\\
  Strategy & 390,525,531  &34,201,922  & 1582,782  &  501,089\\
  \hline
  \end{tabular}
  }
  \label{tb:tb1}
\end{table}

\subsubsection{Evaluation Metrics}
 We evaluate the performance using an offline estimator. 
Through extended engineering, the offline estimator models the user preference and aligns well with the online service. For offline experiments, we evaluate the method with revenue indicator and overestimation indicators. 

\begin{itemize}[leftmargin=*]
  \item \textbf{Reward}. The reward is the total cumulative discount rewards predicted by offline estimator models.
  \item \textbf{Average of Overestimation Degree(AVG-OD)}. The average degree of overestimation on each state-action type. The overestimation degree can be expressed as the difference between the estimated Q value and the actual cumulative rewards.
  
  \begin{equation}
    \text{AVG-OD} = \text{AVG} \Big(Q(s^t,a^t) - \sum_{i=t}^{T} \gamma^{i-t} \cdot r \Big).
\end{equation}
  
  \item \textbf{Standard Deviation of Overestimation Degree(STD-OD)}. The standard deviation of overestimation degree on different state-action type. If the overestimation degree on all state-action types is the same, the STD-OD is small and will not affect the policy's choice of the optimal action. If the overestimation degree is very different and the STD-OD is large, it will seriously harmful the effect of the policy.
  \begin{equation}
     \text{STD-OD} = \text{STD} \Big(Q(s^t,a^t) - \sum_{i=t}^{T} \gamma^{i-t} \cdot r \Big).
\end{equation}
\end{itemize}

\subsubsection{Hyperparameters}

We implement CrossDQN as our baseline model. The learning rate is $10^{-3}$, the optimizer is Adam and the batch size is 8, 192. $\alpha$ is set to 1.0, $\beta$ is set to 0.5 and temperature coefficient is 10.

\subsection{Offline Experiment}

\subsubsection{Baselines} 

In this section, we compare the model performance and the degree of overestimation on following types of data and training strategies:

\begin{itemize}[leftmargin=*]
  \item \textbf{Random Data \& RL}. Only use random data and RL singal.
  \item \textbf{Strategy Data \& RL}. Only use strategy data and RL singal.
  \item \textbf{Strategy Data \& IL}. Only Use strategy data and IL singal.
  \item \textbf{Mixed Data \& RL}. Use strategy and random data and RL singal.
\end{itemize}

\subsubsection{Performance Comparison} 

Table \ref{tb:tb2} summarizes the results of offline experiments. All experiments were repeated 5 times and we have the following observations from the experimental results: i) Due to the unbalanced data distribution, using RL signals to train strategy data or mixed data cannot get good performance on revenue indicator, and overestimation indicators indicate serious overestimation on the trained model. ii) Using RL signals to train random data, the overestimation is mitigated and achieve better performance on revenue indicator. iii) Imposing imitation learning signal to train the strategy data, the estimated Q-value is even smaller than the actual cumulative rewards due to the direct learning of the online policy and the lack of the RL signal. As mentioned above, since the onlien policy used to collect strategy data is trained from random data, its performance on revenue indicator is close to that of using RL signals to train random data. iiii) Our proposed method outperforms all other methods which shows that two kinds of data can be fully utilized to learn better reinforcement learning policy by applying different learning signals. 

 \begin{table}[tb]
      \caption{The results on different types of
data and training strategies. Each experiment are presented in the form of mean ± standard deviation. The improvement means the improvements of MDDL across the best baselines.}
      \renewcommand\arraystretch{1.15}
      \centering
      \setlength{\tabcolsep}{0.84mm}{
        \begin{tabular}{l|ccccc}
          \hline
        Model   & Reward & AVG-OD & STD-OD\\
        \hline
        \hline
        \textbf{Random Data \& RL}  &  2.742($\pm$0.252) & 2.438($\pm$0.271)  & 1.248($\pm$0.238) \\
        \textbf{Strategy Data \& RL} &  2.551($\pm$0.334) & 8.324($\pm$0.346)  & 4.417($\pm$0.295) \\
        \textbf{Strategy Data \& IL} &  2.728($\pm$0.138) & -0.821($\pm$0.122)  & 2.117($\pm$0.156) \\
        \textbf{Mixed Data \& RL} &  2.632($\pm$0.297) & 7.645($\pm$0.327)  & 3.148($\pm$0.285)  \\
        \textbf{MDDL} &  2.887($\pm$0.188) & 2.251($\pm$0.154)  & 1.228($\pm$0.187)  \\
        \hline
        \hline
               \textbf{Improvement} &  5.2\% & -  & -1.6\%  \\
               \hline
        \end{tabular}
      }
      \label{tb:tb2}
 \end{table}

\subsubsection{Hyperparameter Analysis} 

  We keep $\alpha_1$ =1 to analyze the sensitivity of two hyperparameters: $\alpha_2$, $\beta$. 
  Specifically, $\alpha_2$ is the weight of the imitation learning loss and $\beta$ is temperature coefficient on softmax function. i) As $\alpha_2$ increases within a certain range, it's easier to converge to a policy similar to the base model, and to explore a better strategy. When $\alpha_2$ exceeds a certain level, the role played by the exploration data is gradually reduced, and the performance appears to be attenuated to a certain extent. ii) With the increase of $\beta$, the imitation learning signal becomes more and more accurate, which can bring about a certain improvement.

\subsection{Online Results}

  We compare our method with baseline method which all deployed on Meituan food delivery platform through online A/B test. Specifically, we conduct online A/B test with 10\% of whole production traffic from November 09, 2022 to November 15, 2022 (one week). As a result, we gets CTR and GMV increase by 5.46\% and  5.83\% respectively while the overestimation degree is nearly equal. Now, MDDL has been deployed online and serves the main traffic, and contributes to significant business growth.

% \begin{figure}[tb] 
%     \centering 
%     \label{fig:fig3}
%     \subfigure[{The experimental results on the sensitivity of $\alpha_2$.}]{ 
%       \centering
%       \begin{minipage}{\linewidth}
%       \includegraphics[width=\textwidth]{samples/pic7.pdf} 
%       \label{fig:subfig:1} 
%     \end{minipage}%
%     } 
%     \subfigure[{The experimental results on the sensitivity of $\beta$.}]{ 
%       \centering
%       \begin{minipage}{\linewidth}
%       \includegraphics[width=\textwidth]{samples/pic6.pdf} 
%       \label{fig:subfig:2} 
%     \end{minipage}%
%     } 
%     \Description{Experimental results on the sensitivity of $\alpha_2$ and $\beta$.}
%     \caption{{Experimental results on the sensitivity of $\alpha_2$ and $\beta$.}
%     }
%   \end{figure}

% We hope the resulting templates help the reader to write ACM papers
% and proceedings.

\section{Conclusion}

In this paper, we mainly focus on the data aspect which is less researched but equally significant to train RL agent for position allocation problem in feed. We propose MDDL to effectively train RL agents on mixed multi-distribution data by utilizing two kinds of learning signals. For strategy data which is collected from the approximate expert strategy but is highly
imbalanced, we impose the novel imitation
learning signal to reduce the impact of overestimation. For random data which is small scale but relatively uniform, we make full use of the RL signal to enable agent to learn effectively. Practically, both offline experiments and online A/B test have demonstrated the superior performance of our solution.

% This document is important for anybody wanting to comply with the
% requirements of ACM publishing.
\newpage
\balance
\bibliographystyle{ACM-Reference-Format}
\bibliography{abbrev}
\end{document}